\documentclass[
reprint,
amsmath,amssymb,
aps,
]{revtex4-2}

\usepackage{graphicx}% Include figure files
\usepackage{dcolumn}% Align table columns on decimal point
\usepackage{bm}
\usepackage{subcaption}
\usepackage{braket}
\usepackage{ragged2e}
\usepackage{caption}
\usepackage{xcolor}
\usepackage[table]{xcolor}
\usepackage{diagbox}
\usepackage{booktabs}
\usepackage{adjustbox}
\usepackage{multirow}
\usepackage[mathlines]{lineno}
\usepackage{pifont}
\usepackage{makecell}
\usepackage{algorithm}
\usepackage{algpseudocode}
\usepackage{soul}
\usepackage{array}
\usepackage{tikz}
\usetikzlibrary{arrows.meta,calc,positioning}

\newcommand{\boldg}{\mathbf{g}}
\newcommand{\boldb}{\mathbf{b}}

\newcommand{\bolds}{\mathbf{s}}

\newcommand{\boldP}{\mathbf{P}}
\newcommand{\boldQ}{\mathbf{Q}}

\newcommand{\calR}{\mathcal{R}}
\newcommand{\calL}{\mathcal{L}}
\newcommand{\calH}{\mathcal{H}}
\newcommand{\calP}{\mathcal{P}}
\newcommand{\E}{\mathbb{E}}

\newcommand{\C}{\mathbb{C}}
\newcommand{\tr}{\text{tr}}

\newcommand{\cmark}{\textcolor{green!70!black}{\ding{51}}}
\newcommand{\xmark}{\textcolor{red}{\ding{55}}}
\newcommand{\hstab}{\hspace*{1.5em}}

\usepackage{amsthm}
\newtheorem{remark}{Remark}

\usepackage{bibunits}
\defaultbibliographystyle{apsrev4-2}
\begin{document}

% SUPP REF
% \begin{bibunit}

% \preprint{APS/123-QED}

\title{\textbf{Verifying random matrix product states with autoregressive local measurements}}

% Author block\author{Hyunho Cha}
% \author{\textcolor{red}{Author}}
% \email{\textcolor{red}{EMAIL}}
% \affiliation{\textcolor{red}{Affiliation}}
\author{Hyunho Cha}
\email{ovalavo@snu.ac.kr}
\author{Subin Kim}
\email{subini0213@snu.ac.kr}
\author{Jungwoo Lee}
\email{junglee@snu.ac.kr}
\affiliation{NextQuantum and Department of Electrical and Computer Engineering, Seoul National University, Seoul 08826, Republic of Korea}

\date{\today}% It is always \today, today,
             %  but any date may be explicitly specified

\begin{abstract}
Matrix product states (MPS) are a central language for one-dimensional quantum matter and a practical target for near-term quantum simulators and variational algorithms.
Yet, while substantial effort has focused on preparing MPS with shallow circuits, scalable methods to \emph{verify} that a many-body device has actually produced the intended state remain underdeveloped.
Direct fidelity estimation (DFE) relies only on local Pauli measurements, but in many-body settings it suffers an exponential classical overhead from the preprocessing needed to sample Pauli strings.
We eliminate this obstacle by introducing an \emph{autoregressive} importance sampler that draws Pauli strings sequentially from efficiently computable conditional distributions, reducing the per-shot classical overhead to linear scaling in the number of qubits.
We further develop a grouped extension that constructs qubit-wise commuting measurement settings via a \emph{sorting string} and simultaneously estimates the entire commuting group from a single setting, significantly reducing estimator variance while preserving efficient postprocessing.
Our approach extends naturally to matrix product operators (MPO), enabling scalable verification of tensor-network states and observables in long one-dimensional quantum systems.
We utilize random MPS as a natural benchmark for generic 1D entangled states.
\end{abstract}

%\keywords{Suggested keywords}%Use showkeys class option if keyword
                              %display desired
\maketitle

%\tableofcontents

\section{Introduction}

\begin{figure}[t]
    \centering
    \includegraphics[width=\linewidth]{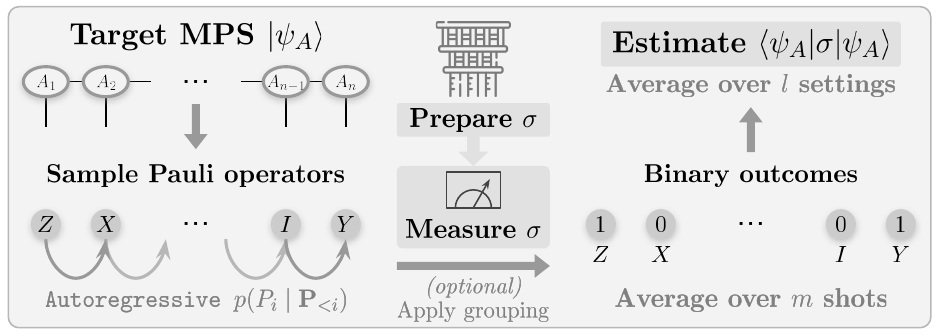}
    \caption{\justifying Schematic of our MPS-(G)DFE protocol. Given an experimental state $\sigma$ and a target MPS $|\psi_A\rangle$, we estimate the fidelity $\langle\psi_A|\sigma|\psi_A\rangle$ using only local Pauli measurements. We sample Pauli strings $\boldP$ sequentially and autoregressively from the target weight distribution $p(\boldP)$, measure $\sigma$ in the sampled setting $\boldP$ to estimate $\langle\boldP\rangle_\sigma$, and aggregate the resulting single-shot estimators.}
    \label{fig:schematic}
\end{figure}

Quantum state verification is a core capability of quantum information science \cite{takeuchi2018verification, eisert2020quantum, kliesch2021theory, yu2022statistical}. To benchmark hardware, validate quantum simulations, and assess whether a device prepares the intended entangled state, one must reliably estimate figures of merit such as the fidelity between an experimental state and an ideal target \cite{pallister2018optimal, zhu2019efficient, zambrano2024certification}.
Yet the brute-force route (full quantum state tomography) quickly becomes infeasible as system size grows, with measurement and classical post-processing costs that scale exponentially in the number of qubits \cite{cramer2010efficient, flammia2010heralded, gross2010quantum, eisert2020quantum, wang2020scalable, kliesch2021theory}.
This scalability barrier is especially acute for many-body states produced by near-term processors and analog simulators, where we often have a concrete target in mind but cannot afford to reconstruct the entire density matrix.

Direct fidelity estimation (DFE) provides an attractive alternative to tomography \cite{da2011practical, flammia2011direct}.
Rather than learning the full state, DFE estimates the fidelity itself by measuring a randomized subset of Pauli observables, sampled according to an importance-weighted distribution determined by the known target state.
In favorable cases, the number of state preparations needed to reach a desired precision does not depend on the Hilbert-space dimension, offering an exponential advantage in principle.
However, the practical efficiency of standard DFE can break down for generic highly entangled targets \cite{cerezo2020variational, zhang2021direct, leone2023nonstabilizerness}.
Their Pauli expansions typically spread weight over exponentially many strings, making it costly to compute or sample from the importance distribution, and potentially inflating estimator variance when small Pauli coefficients appear.
Related randomized-measurement approaches including \emph{classical shadows} have significantly broadened the verification toolkit, but a broadly applicable method that exploits structure while remaining compatible with local measurements is still highly desirable \cite{huang2020predicting, elben2023randomized, gupta2025few, huang2025certifying, coladangelo2026power}.

Many experimentally relevant many-body states are structured rather than arbitrary \cite{fannes1992finitely, vidal2003efficient, eisert2010colloquium, schollwock2011density, orus2014practical, cirac2021matrix}.
In one dimension, matrix product states (MPS) provide an efficient representation for states with limited entanglement, controlled by a bond dimension $B_\mathrm{s}$ \cite{white1992density, perez2006matrix, vidal2007classical, verstraete2008matrix}.
While prior works have focused on the efficient preparation of MPS \cite{schon2005sequential, schon2007sequential, smith2024constant}, methods for verifying successful preparation remain underdeveloped.
MPS underlie powerful classical methods such as DMRG and enable efficient computation of local observables and measurement statistics via tensor contractions.
This observation suggests an opportunity for verification.
When the target is given as an MPS, we can use that description to design measurements without ever enumerating exponentially many Pauli terms.

Estimating the fidelity to a pure target state using random Clifford measurements generally yields better sample complexity than using random Pauli measurements \cite{huang2020predicting, akhtar2023scalable, bertoni2024shallow}. However, Clifford circuits are often hardware-inefficient to implement. Furthermore, this approach requires the evaluation of $|\langle\boldb|U|\psi\rangle|^2$ for Clifford $U$, computational-basis $\boldb$, and MPS $|\psi\rangle$. This task is \#P-hard already for product-state inputs (hence already for bond-dimension-1 MPS) \cite[Theorem 6]{jozsa2013classical}.

Random MPS furnish a principled benchmark class inside the MPS manifold \cite{garnerone2010typicality, lancien2022correlation, haag2023typical, chen2024magic, jameson2024optimal, malz2024preparation, leontica2025unbiased, jaderberg2026variational}.
In this work, we introduce an efficient protocol for fidelity estimation tailored to random MPS targets using \emph{autoregressive} sampling of Pauli strings.
Instead of choosing Pauli bases from a global distribution, we sample the $n$-qubit measurement setting sequentially along the chain.
The known MPS is used to compute conditional distributions for the next local basis given previously chosen bases.
In this way we draw full Pauli strings directly from the target's native importance distribution with only \emph{linear} overhead in system size, eliminating the exponential preprocessing bottleneck of naive DFE for generic states.

We further develop a grouped extension that partitions sampled Pauli strings into qubit-wise commuting groups so that multiple observables can be inferred from a single measurement setting.
This grouping reduces experimental runs and can substantially suppress estimator variance, while requiring only coordinated single-qubit basis choices.

Finally, the approach extends naturally from MPS to matrix product operators (MPO), enabling scalable observable estimation for structured targets and one-dimensional processes with analogous contraction costs \cite{baumgratz2013scalable, guo2024quantum, qin2024quantum}.
Overall, our results realize the original promise of DFE across a broad and practically important class of many-body targets by explicitly leveraging tensor-network structure.

\section{Preliminaries}

\subsection{Basic notation}

We consider an $n$-qubit system with Hilbert space
$\mathcal{H}_d := (\mathbb{C}^2)^{\otimes n}$ of dimension $d = 2^n$.
We write $\mathcal{L}(\mathcal{H}_d)$ for the space of linear operators on $\mathcal{H}_d$.

Let $\calP_1 := \{I, X, Y, Z\}$ denote the set of single-qubit Pauli matrices, and let
\[
\calP_n := \calP_1^{\otimes n}
= \bigl\{ P_1 \otimes \cdots \otimes P_n \;:\; P_i \in \calP_1 \bigr\}
\]
denote the set of $n$-qubit Pauli strings.
We will sometimes represent a Pauli string by its \emph{label vector}
$\boldP = (P_1,\dots,P_n) \in \calP_1^n$.
By a mild abuse of notation, we also write $\boldP$ for the corresponding operator
$\bigotimes_{i=1}^n P_i \in \calP_n$; the intended meaning will be clear from context.

\subsection{Matrix product states}

MPS are a special class of tensor networks (see Appendix~\ref{sec:tensor_networks} for a brief overview of graphical notation and tensor contraction).

An \emph{open-boundary} $n$-qubit MPS is specified by a collection of local tensors
\[
A = (A_1,\dots,A_n),
\]
together with bond dimensions $\alpha_0,\dots,\alpha_n$ satisfying $\alpha_0=\alpha_n=1$.
Each local tensor has two bond indices and one physical index:
\[
A_i \in \C^{\alpha_{i-1}\times \alpha_i \times 2}\quad (i=1,\dots,n),
\]
where the last dimension $2$ corresponds to the single-qubit computational basis $\{0,1\}$.
The corresponding (pure) state vector in $\mathcal{H}_d$ is
\[
|\psi_A\rangle := \sum_{b\in\{0,1\}^n} \sum_{i_0, \dots , i_n} A_{1, i_0, i_1, b_1} \times \cdots \times A_{n, i_{n-1}, i_n, b_n} |b\rangle.
\]
Throughout, we assume $\ket{\psi_A}$ is normalized. Also, we write
\[
\rho_A := |\psi_A\rangle\langle\psi_A|
\]
for the associated density operator.

\subsection{Direct fidelity estimation}
The DFE method, as proposed by \cite{flammia2011direct}, leverages the sampling of Pauli operators to efficiently estimate the fidelity of a quantum state. The process is described by the mathematical framework below.
\paragraph{The characteristic function.}
To begin, we represent any Hermitian operator $A \in \mathcal{L}(\mathcal{H}_d)$ through its characteristic function. For a given Pauli operator $\mathbf{P}$, this is defined as the normalized trace:
$$
\chi_{A,\mathbf{P}} = \frac{\tr(A \mathbf{P})}{\sqrt{d}}.
$$
Essentially, this value represents the scaled expectation value of $\mathbf{P}$ with respect to the operator $A$.
\paragraph{Fidelity via Pauli decomposition.}
Consider an unknown quantum state $\sigma$ and a target pure state $\rho$. The fidelity between these two states, given by $\tr(\rho\sigma)$, can be decomposed into the sum of their overlapping characteristic functions:
$$
\tr(\rho\sigma) = \sum_\mathbf{P} \chi_{\rho,\mathbf{P}} \chi_{\sigma,\mathbf{P}}.
$$
\paragraph{The estimation strategy.}
Rather than measuring every Pauli operator, we use an importance sampling approach. We treat the squared characteristic functions of the target state as a probability distribution. Since $\rho$ is a pure state, the values $p(\mathbf{P}) = \chi_{\rho,\mathbf{P}}^2$ are guaranteed to sum to one, forming a valid distribution for selecting $\mathbf{P}$ at random. Given precision parameters $(\epsilon,\delta)$, we sample $l=\lceil1/\epsilon^2\delta\rceil$ independent random variables $R_\boldP$ associated with the chosen Pauli operators:
$$
R_\boldP := \frac{\chi_{\sigma,\mathbf{P}}}{\chi_{\rho,\mathbf{P}}}.
$$
Taking the expected value of this variable yields the fidelity, i.e., $\E_\boldP[R_\boldP] = \tr(\rho\sigma)$. In practice, because the characteristic function of the unknown state ($\chi_{\sigma,\mathbf{P}}$) is not known beforehand, it must be determined statistically through repeated experimental measurements of the sampled Pauli operators.
The number of shots required for each sampled $\mathbf{P}$ is given by
$$
m = \left\lceil \frac{2}{p(\boldP)dl\epsilon^2} \ln\frac{2}{\delta} \right\rceil.
$$

\paragraph{Guarantee.}
With an appropriate choice of the number of sampled settings $l$ and the number of shots per setting (used to estimate each $\chi_{\sigma,\boldP}$),
the resulting estimator achieves additive error at most $\epsilon$ with failure probability at most $\delta$, i.e.,
\[
p\!\left(\left|\widehat{\tr(\rho\sigma)}-\tr(\rho\sigma)\right|\le \epsilon\right)\ge 1-\delta.
\]

\subsection{Grouping Pauli operators}
To further reduce estimator variance, we adopt \emph{grouped} direct fidelity estimation (GDFE), which partitions the Pauli strings into \emph{qubit-wise commuting} (QWC) groups~\cite{verteletskyi2020measurement, crawford2021efficient, hadfield2022measurements, barbera2025sampling, dalfavero2025measurement}. Each QWC group corresponds to a single measurement setting, so one setting yields unbiased estimates for all Pauli strings in the group. In GDFE, groups are sampled according to their total weight, and repeated measurements within the sampled setting are used to estimate the grouped contribution.
Full details are provided in Appendix~\ref{subsec:grouping-pauli}.

\section{Main results}

Applying direct fidelity estimation (DFE) to highly structured target states such as GHZ and W is comparatively straightforward, because the Pauli-weight distribution and the corresponding characteristic values admit explicit expressions, so one can efficiently sample Pauli strings and evaluate the required weights \cite{flammia2011direct, li2020optimal}.
For more generic many-body targets, this preprocessing quickly becomes intractable, since it would require brute-force evaluation of the characteristic values for all $d^2$ Pauli strings.

We show that this bottleneck disappears for MPS targets by developing an efficient autoregressive sampling algorithm. Our goal is as follows. Given an unknown prepared state $\sigma$ and a target MPS $|\psi_A\rangle$ with density operator $\rho_A = |\psi_A\rangle\langle\psi_A|$, we estimate the fidelity
\[
\langle \psi_A|\sigma|\psi_A\rangle = \tr(\rho_A \sigma)
\]
using only local Pauli measurements on independent copies of $\sigma$.

\subsection{MPS-DFE with autoregressive local measurements}
\label{section:main_algorithm}

To implement DFE with an MPS target, we must sample Pauli strings $\boldP$ from the distribution $\chi_{\rho_A,\boldP}^2$ and evaluate $\chi_{\rho_A,\boldP}$ for each sampled $\boldP$.

A brute-force evaluation of $\chi_{\rho_A,\boldP}$ over all $4^n$ Pauli strings is intractable. Instead, we sample $\boldP$ autoregressively, drawing one single-qubit Pauli at a time. Write
\[
\boldP = (P_1, \dots , P_n).
\]
At each step $1\le i\le n$, we sample \(P_i\) from the conditional distribution
\[
p(P_i \mid \boldP_{<i}).
\]
To compute these conditional probabilities efficiently, we introduce a convenient ``pairing'' permutation. Define a $2n$-qubit permutation operator $S_n\in\mathrm{SU}(2^{2n})$ such that
$$
S_n|b,b^\prime\rangle = |b_1,b^\prime_1,\dots,b_n,b^\prime_n\rangle
$$
for all \(b,b^\prime \in \{0,1\}^n\), and set
$$
\tilde\rho^{(2)} := S_n \rho_A^{\otimes2} S_n^\dagger.
$$
The key observation is that the sampling weight can be written as
\begin{align}
\label{equation:prob_as_double_copy}
p(\boldP)
& = \chi_{\rho_A,\boldP}^2\nonumber\\
& = \frac{1}{d} \tr(\rho_A \boldP)^2\nonumber\\
& = \frac{1}{d} \tr\!\left((\rho_A \boldP)^{\otimes2}\right)\\
& = \frac{1}{d} \tr\!\left(\rho_A^{\otimes2} \boldP^{\otimes2}\right)\nonumber\\
& = \frac{1}{d} \tr\!\left( \tilde\rho^{(2)} \bigotimes_{i=1}^n P_i^{\otimes2} \right).\nonumber
\end{align}
Moreover,
\begin{align}
\label{equation:summation_as_swap}
\sum_{\boldP\in\calP_n} \boldP^{\otimes2} & = S_n^\dagger \left(\sum_{\boldP\in\calP_1^n} \bigotimes_{i=1}^n P_i^{\otimes2}\right) S_n\nonumber\\
& = S_n^\dagger \left(\sum_{P\in\calP_1} P^{\otimes2}\right)^{\otimes n} S_n\\
& = S_n^\dagger (2\times \mathrm{SWAP})^{\otimes n} S_n.\nonumber
\end{align}
Using these identities, we can efficiently evaluate the partial sums that appear in conditional probabilities (see Appendix~\ref{section:detailed_conditional_swap_equation}).

Since
\[
p(\boldP) = \prod_{i=1}^n p(P_i \mid \boldP_{<i}),
\]
we can sample $\boldP$ sequentially using
\begin{equation}
\label{eq:conditional_formula}
p\!\left(P_i \,\middle|\, \boldP_{<i}=\boldP^{(\mathrm{l})}\right) = \frac{\sum_{\boldP_{\le i}=\left(\boldP^{(\mathrm{l})};P_i\right)} p(\boldP)}{\sum_{\boldP_{<i}=\boldP^{(\mathrm{l})}} p(\boldP)}.
\end{equation}

\begin{figure*}
  \centering
  \begin{subfigure}[c]{0.63\linewidth}
    \centering
    \includegraphics[width=\linewidth]{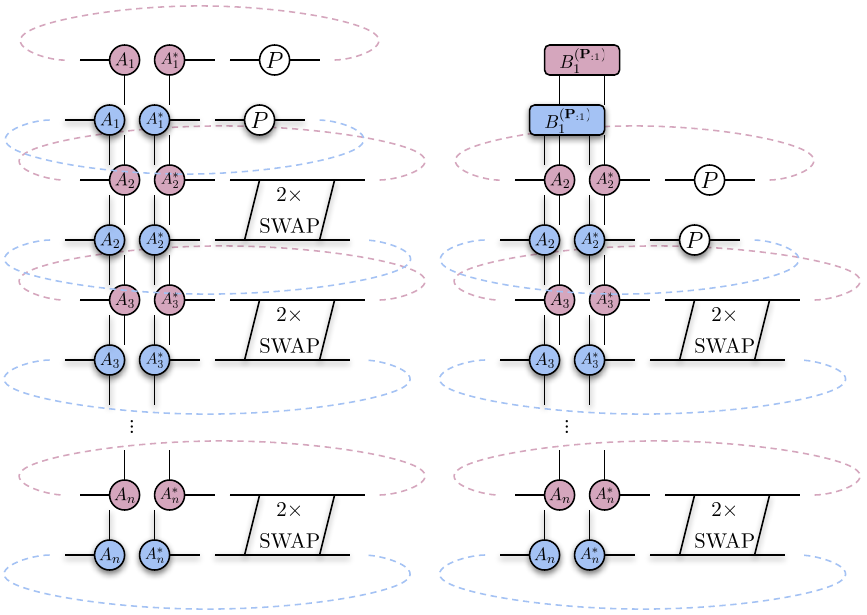}
    \caption{}
    \label{fig:TN_a}
  \end{subfigure}
  \begin{subfigure}[c]{\linewidth}
    \centering
    \includegraphics[width=0.85\linewidth]{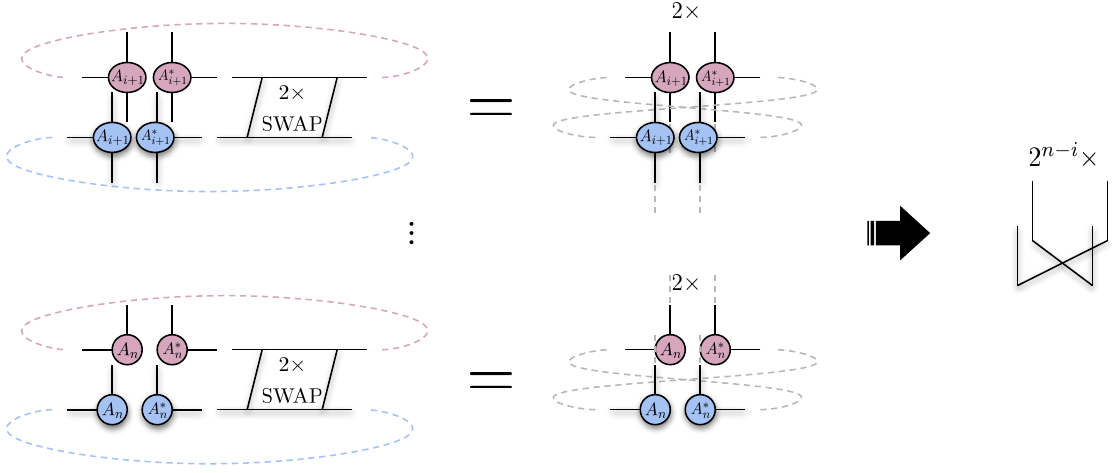}
    \caption{}
    \label{fig:swap_simpler}
  \end{subfigure}
  \caption{\justifying (a) The first two steps of our autoregressive sampling procedure. Dashed lines indicate contractions. Here $A_\cdot^\ast$ denotes the complex conjugate of $A_\cdot$. (b) If the MPS is in right-canonical form with orthogonality center at the first tensor, then each backward SWAP-contraction reduces to the crossed identity.}
  \label{fig:contraction_example}
\end{figure*}

% comment block 1

We now describe the resulting algorithm in detail. Without loss of generality, we first gauge-transform the target MPS into right-canonical form with orthogonality center at the first tensor (see Appendix~\ref{sec:tensor_networks}). Define the local tensors
\begin{equation}
\label{eq:B_tilde_definition}
\tilde{B}^{(P)}_{i,a,b,c,d}
=
\sum_{e,f}
A_{i,a,c,e}A^{*}_{i,b,d,f}P_{f,e}
\end{equation}
for \(1\le i\le n\) and \(P\in\calP_1\).

\begin{remark}
The tensor $\tilde B$ is introduced for notational convenience rather than computational efficiency. In fact, $\tilde B$ is always interpreted as the right-hand side of Eq.~\eqref{eq:B_tilde_definition} in subsequent steps, where we contract $A$ and $A^\ast$ separately.
\end{remark}

In this gauge, once the Pauli operators on the unsampled suffix are summed using Eq.~\eqref{equation:summation_as_swap}, each local $\mathrm{SWAP}$ simply exchanges the two tensor pieces, and the contraction of the remaining right-canonical suffix with its conjugate reduces to the \emph{crossed identity} on the incoming doubled bond (see Fig.~\ref{fig:swap_simpler}). Hence no separate family of backward $\mathrm{SWAP}$-contraction tensors needs to be precomputed.

Sampling now proceeds by a single forward sweep. Initialize
\[
B^{(\emptyset)}_{0}:=1.
\]
For $1\le i\le n$, suppose that $\boldP_{:(i-1)}$ has already been sampled and that the corresponding forward message $B^{\left(\boldP_{:(i-1)}\right)}_{i-1}$ has been constructed. For each candidate $P\in \calP_1$, define
\[
\hat{B}^{(P)}_{i,a,b}
:=
\sum_{c,d}
B^{\left(\boldP_{:(i-1)}\right)}_{i-1,c,d}\,
\tilde{B}^{(P)}_{i,c,d,a,b}.
\]
Then
\[
\sum_{\boldP_{>i}\in \calP_1^{n-i}}
p\bigl(\boldP_{:(i-1)}, P, \boldP_{>i}\bigr)
=
\frac{2^{n-i}}{d}
\sum_{a,b}
\hat{B}^{(P)}_{i,a,b}\hat{B}^{(P)}_{i,b,a}.
\]
Since the prefactor $2^{\,n-i}/d$ is independent of $P$, we define the unnormalized conditional weights
\[
\beta_i^{(P)}
:=
\sum_{a,b}
\hat{B}^{(P)}_{i,a,b}\hat{B}^{(P)}_{i,b,a}
=
\operatorname{tr}\!\left[\bigl(\hat{B}^{(P)}_{i}\bigr)^2\right].
\]
Because $\hat{B}^{(P)}_{i}$ is Hermitian, $\beta_i^{(P)}\ge 0$, and therefore
\[
\frac{\beta_i^{(P)}}{\sum_{P'\in \calP_1}\beta_i^{(P')}}
=
p(P_i=P\mid \boldP_{<i}).
\]
We thus sample $P_i$ according to these weights and then set
\[
B_i^{(\boldP_{:i})}:=\hat{B}^{(P_i)}_{i}.
\]

Repeating this update for $i=1,\dots,n$ yields a full Pauli string $\boldP$ sampled from $p(\boldP)$. After the final step, $B_n^{(\boldP_{:n})}=B_n^{(\boldP)}$ is a scalar and satisfies
\[
B^{(\boldP)}_{n,1,1}
=
\operatorname{tr}(\rho_A \boldP).
\]
Hence
\[
p(\boldP)=\frac{1}{d}\bigl(B^{(\boldP)}_{n,1,1}\bigr)^2,
\qquad
\chi_{\rho_A,\boldP}
=
\frac{1}{\sqrt d}\,B^{(\boldP)}_{n,1,1}.
\]
The first two sampling steps are illustrated in Fig.~\ref{fig:TN_a}.

\subsection{MPS-GDFE with autoregressive local measurements}
\label{sec:mps_gdfe_derivation}

We sample $\boldP$ exactly as in MPS-DFE. Here, however, we interpret $\boldP$ as a \emph{latent} Pauli string that is associated with a \emph{group} of Pauli operators that can be estimated from a single measurement setting.

We begin by specifying a grouping rule. For the $n$-qubit positions, we first draw a \emph{sorting string}
$$
\boldg \in \{X,Y,Z\}^n.
$$
Intuitively, $\boldg$ designates, at each qubit $i$, a preferred non-identity Pauli label $g_i \in \{X,Y,Z\}$ that will be used to form a representative measurement setting. Now suppose we have sampled a latent Pauli string $\boldP \in \calP_1^n$ using the method described in Section~\ref{section:main_algorithm}. Given $\boldg$, we define the \emph{representative} Pauli string associated with $\boldP$ as $\calR_\boldg(\boldP)$, where
$$
\calR_\boldg(\boldP)_i = \begin{cases}
    g_i & \text{if } P_i \in \{I,g_i\}\\
    P_i & \text{otherwise}
\end{cases}.
$$
The representative string $\calR_\boldg(\boldP)$ naturally induces a group $g(\calR_\boldg(\boldP))$ of latent Pauli strings that share the same representative, where $g(\cdot)$ is defined by
$$
g(\boldQ) := \big\{ \boldP \in \calP_1^n \mid \calR_\boldg(\boldP)=\boldQ \big\}
$$
for all $\boldQ \in \{X,Y,Z\}^n$. Our goal is to estimate, from a single measurement setting, all Pauli operators in the group $g(\calR_\boldg(\boldP))$ simultaneously. Define
$$
p_\boldg(\boldP) := \sum_{\boldP^\prime \in g(\calR_\boldg(\boldP))} \chi_{\rho_A, \boldP^\prime}^2.
$$
From Eq.~\eqref{eq:group_rv_ideal} in Appendix~\ref{subsec:grouping-pauli}, our target is the random variable
$$
R_\boldP := \frac{\sum_{\boldP^\prime \in g(\calR_\boldg(\boldP))} \chi_{\rho_A, \boldP^\prime} \chi_{\sigma, \boldP^\prime}}{p_\boldg(\boldP)}.
$$
Suppose we measure $\sigma$ in the setting $\calR_\boldg(\boldP)$ and obtain signs $\bolds \in \{\pm1\}^n$. Define
$$
M^{(\boldP,\bolds)}_i := \begin{cases}
    2\Pi_{s_i}^{(g_i)} & \text{if } P_i \in \{I,g_i\}\\
    s_i P_i & \text{otherwise}
\end{cases},
$$
where $\Pi_{s_i}^{(g_i)}$ is the rank-1 projector onto the $s_i$-eigenspace of $g_i$. We then define a single-shot snapshot estimator of $R_\boldP$ as
\begin{equation}
\label{equation:snapshot_definition}
\tilde R_{\boldP,\bolds} = \frac{1}{p_\boldg(\boldP)d} \tr\!\left(\rho_A \bigotimes_{i=1}^n M^{(\boldP,\bolds)}_i \right).
\end{equation}
One can verify that this estimator is unbiased, i.e., $\E_\bolds\!\left[\tilde R_{\boldP,\bolds}\right] = R_\boldP$ (see Appendix~\ref{section:mps_gdfe_snapshot_unbiased}).

\begin{table}[t]
\caption{\justifying 5-qubit illustration of the grouping and snapshot construction. Given a sorting string $\boldg$ and a sampled latent Pauli string $\boldP$, we form the representative $\calR_\boldg(\boldP)$ and its associated group $g(\calR_\boldg(\boldP))$. Measuring $\sigma$ in the setting $\calR_\boldg(\boldP)$ yields a bitstring, which determines the local factors $M^{(\boldP,\bolds)}_i$ used in the snapshot estimator.}
\label{table:grouping_example}
\centering
\renewcommand{\arraystretch}{1.5}
% \begin{tabular}{|c|c|c|c|c|}
\begin{tabular}{!{\vrule width 1pt}c|c|c|c|c!{\vrule width 1pt}}
\Xhline{1pt}
\multicolumn{5}{!{\vrule width 1pt}c!{\vrule width 1pt}}{\cellcolor{gray!10}\textbf{Sorting string} $\boldg$}\\
\hline
\hstab$Z$\hstab & \hstab$X$\hstab & \hstab$Y$\hstab & \hstab$Z$\hstab & \hstab$X$\hstab\\
\Xhline{1pt}
\multicolumn{5}{!{\vrule width 1pt}c!{\vrule width 1pt}}{\cellcolor{gray!10}\textbf{Sampled} $\boldP$}\\
\hline
$X$ & $I$ & $Y$ & $I$ & $Z$\\
\Xhline{1pt}
\multicolumn{5}{!{\vrule width 1pt}c!{\vrule width 1pt}}{\cellcolor{gray!10}$\calR_\boldg(\boldP)$}\\
\hline
$X$ & $X$ & $Y$ & $Z$ & $Z$\\
\Xhline{1pt}
\multicolumn{5}{!{\vrule width 1pt}c!{\vrule width 1pt}}{\cellcolor{gray!10}$g(\calR_\boldg(\boldP))$}\\
\hline
$X$ & $I/X$ & $I/Y$ & $I/Z$ & $Z$\\
\Xhline{1pt}
\multicolumn{5}{!{\vrule width 1pt}c!{\vrule width 1pt}}{\cellcolor{gray!10}\textbf{Measured signs} $\bolds$}\\
\hline
$-1$ & $+1$ & $+1$ & $-1$ & $+1$\\
\Xhline{1pt}
\multicolumn{5}{!{\vrule width 1pt}c!{\vrule width 1pt}}{\cellcolor{gray!10}\textbf{Factorization} $M^{(\boldP,\bolds)}_i$}\\
\hline
$-X$ & $I+X$ & $I+Y$ & $I-Z$ & $+Z$\\
\Xhline{1pt}
\end{tabular}
\end{table}

Table~\ref{table:grouping_example} provides a step-by-step 5-qubit illustration of the grouping procedure and snapshot construction.

From Eq.~\eqref{eq:shots_per_setting} in Appendix~\ref{subsec:grouping-pauli}, the number of snapshots per measurement setting should be set to
$$
m=\left\lceil \frac{2 \left(\sum_{\boldP^\prime \in g(\calR_\boldg(\boldP))} \left|\chi_{\rho_A, \boldP^\prime}\right|\right)^2}{p_\boldg(\boldP)^2 d l \epsilon^2} \ln\!\frac{2}{\delta} \right\rceil,
$$
but the term $\left(\sum_{\boldP^\prime \in g(\calR_\boldg(\boldP))} \left|\chi_{\rho_A, \boldP^\prime}\right|\right)^2$ cannot be evaluated in polynomial time. Instead, we can safely upper bound it by $p_\boldg(\boldP) |g(\calR_\boldg(\boldP))|$ (via Cauchy–Schwarz), which preserves the bounds in \cite{barbera2025sampling} showing that GDFE improves over DFE in both estimator variance and total number of shots, up to the usual looseness incurred by this inequality. With this substitution, we set
\begin{equation}
\label{eq:l2_version_num_copies}
\tilde m =\left\lceil \frac{2 |g(\calR_\boldg(\boldP))|}{p_\boldg(\boldP) d l \epsilon^2} \ln\!\frac{2}{\delta} \right\rceil.
\end{equation}
This replacement is useful because both $|g(\calR_\boldg(\boldP))|$ and $p_\boldg(\boldP)$ can be computed efficiently on the fly as we sample $\boldP$. First,
$$
|g(\calR_\boldg(\boldP))| = 2^{|\{i:\calR_\boldg(\boldP)_i = g_i\}|}.
$$
Next, we describe how to compute $p_\boldg(\boldP)$. The procedure is conceptually the same as the sequential contraction used to compute the characteristic function in MPS-DFE (illustrated in Fig.~\ref{fig:TN_a}), except that at certain qubit sites we must sum over \emph{two} possible single-qubit Pauli operators. For $a,Q \in \{X,Y,Z\}$, define
\begin{equation}
\label{eq:preimage_definition}
r^{-1}_a(Q) := \begin{cases}
    \{I,Q\} & \text{if } Q=a\\
    \{Q\} & \text{otherwise}
\end{cases},
\end{equation}
i.e., the single-qubit preimage of $Q$ under the representative map induced by $a$. We initialize
$$
G^{(\boldg,\boldP)}_{1,a,b,c,d}
% = \sum_{P \in r^{-1}_{a_1}(\calR_\boldg(\boldP)_1)} \sum_{e,f,g,h} A_{1,1,a,e} A^\ast_{1,1,b,f} A_{1,1,c,g} A^\ast_{1,1,d,h} P_{f,e} P_{h,g}.
= \sum_{P \in r^{-1}_{g_1}(\calR_\boldg(\boldP)_1)} \tilde B^{(P)}_{1, 1, 1, a, b} \tilde B^{(P)}_{1, 1, 1, c, d}.
$$
Then, for $2\le i\le n$, immediately after we sample $P_i$, we compute
% \begin{widetext}
\begin{align*}
& G^{(\boldg,\boldP)}_{i,a,b,c,d}\\
% = \sum_{P \in r^{-1}_{a_{i+1}}(\calR_\boldg(\boldP)_{i+1})} \sum_{e,f,g,h,i^\prime,j,k,l} G_{i,e,f,g,h} A_{i+1,e,a,i^\prime} A^\ast_{i+1,f,b,j} A_{i+1,g,c,k} A^\ast_{i+1,h,d,l} P_{j,i^\prime} P_{l,k}.
& = \sum_{P \in r^{-1}_{g_{i}}(\calR_\boldg(\boldP)_{i})} \sum_{e,f,g,h} G^{(\boldg,\boldP)}_{i-1,e,f,g,h} \tilde B^{(P)}_{i, e, f, a, b} \tilde B^{(P)}_{i, g, h, c, d}.
\end{align*}
% \end{widetext}
By construction, this yields
$$
G^{(\boldg,\boldP)}_{n,1,1,1,1} = \sum_{\boldP \in g(\calR_\boldg(\boldP))} \tr(\rho_A \boldP)^2,
$$
and therefore
$$
p_\boldg(\boldP) = \frac{1}{d} G^{(\boldg,\boldP)}_{n,1,1,1,1}.
$$

\begin{remark}
The sorting string $\boldg \in \{X,Y,Z\}^n$ may be treated either as a deterministic auxiliary choice or as an additional source of randomness in the grouped protocol. In particular, one may either fix $\boldg$ once and use that same string throughout the full estimation procedure, or redraw $\boldg$ independently for each sampled latent Pauli string $\boldP$.

The unbiasedness argument in Appendix~\ref{section:mps_gdfe_snapshot_unbiased} is conditional on the realized value of $\boldg$. Therefore, for every fixed $\boldg$, the snapshot estimator is unbiased for the corresponding grouped random variable. Consequently, if $\boldg$ is itself randomized, the estimator remains unbiased after averaging over the additional randomness as well.

When one wishes to randomize $\boldg$, a natural default choice is the uniform distribution over $\{X,Y,Z\}^n$. In our numerical experiments, each run begins by drawing one $\boldg$ uniformly at random from $\{X,Y,Z\}^n$, which is then held fixed throughout that run.

At present, we do not claim that one implementation is uniformly preferable. Comparing these choices at the level of finite-sample efficiency depends on how the group weights and shot allocations vary with $\boldg$. Such comparisons should be interpreted at the level of the practical shot-allocation rule in Eq.~\eqref{eq:l2_version_num_copies}.
A more systematic study of how the choice of $\boldg$ affects performance is left for future work.
\end{remark}

\subsection{MPO targets}
\label{sec:mpo_algorithm}

While our primary focus is on MPS targets, this approach extends naturally to \emph{Hermitian} MPO targets.
Explicit normalization is essential here, as the characteristic function generally does not have a unit norm.

An open-boundary $n$-qubit matrix product operator (MPO) is described by
$n$ tensors $\tilde A = (\tilde A_1,\ldots,\tilde A_n)$ with bond dimensions
$\beta_0,\ldots,\beta_n$ where $\beta_0=\beta_n=1$.
Each local tensor has two physical indices (input/output) and two bond indices:
\[
\tilde A_i \in \mathbb{C}^{\beta_{i-1}\times \beta_i \times 2 \times 2}.
\]
The operator corresponding to $\tilde A$ is denoted
$O_{\tilde A}\in\calL(\calH_d)$ and defined by
\begin{align*}
O_{\tilde A}
=
\sum_{b,b^\prime\in\{0,1\}^n}
\sum_{i_0,\ldots,i_n}
& \tilde A_{1,i_0,i_1,b_1,b^\prime_1}
\times\\
\cdots \times
& \tilde A_{n,i_{n-1},i_n,b_n,b^\prime_n}
|b\rangle\langle b^\prime|.
\end{align*}
For an MPO \(\tilde A\), we define
$$
Z
=\sum_{\boldP\in\calP_1^{\otimes n}}\chi_{O_{\tilde A},\boldP}^2
=\sum_{\boldP\in\calP_1^{\otimes n}} \frac{1}{d} \tr(O_{\tilde A} \boldP)^2
$$
and we sample $\boldP$ from the distribution \(\chi_{O_{\tilde A},\boldP}^2/Z\). Note that \(Z\) can be computed efficiently using the procedure described in Section~\ref{section:main_algorithm}. With this normalization, the random variable associated with $\boldP$ should be defined as \(Z\chi_{\sigma,\boldP}/\chi_{O_{\tilde A},\boldP}\).

For $1\le i\le n$ and $P\in \calP_1$, define the operator-space local matrices
\[
\Gamma_i^{(P)} \in \mathbb{C}^{\beta_{i-1}\times \beta_i},
\qquad
\Gamma_{i,a,b}^{(P)}
:=
\frac{1}{\sqrt{2}}
\sum_{e,f=0}^1
\tilde{A}_{i,a,b,e,f}\,P_{f,e}.
\]
Then
\[
\chi_{O_{\tilde{A}},\boldP}
=
\left(\Gamma_1^{(P_1)}\Gamma_2^{(P_2)}\cdots \Gamma_n^{(P_n)}\right)_{1,1}.
\]
Thus the MPO induces an MPS of physical dimension $2^2=4$ on the single-qubit Pauli alphabet $\calP_1$.

Without loss of generality, we first gauge-transform this induced MPS into right-canonical form with orthogonality center at the first tensor, i.e.,
\[
\sum_{P\in \calP_1}\Gamma_i^{(P)}\Gamma_i^{(P)\dagger}
=
I_{\beta_{i-1}},
\qquad
i=2,\ldots,n.
\]
Sampling now proceeds by a single forward sweep. Initialize
\[
v_0^{(\emptyset)}:=1.
\]
For $1\le i\le n$, suppose that $\boldP_{:(i-1)}$ has already been sampled and that the corresponding forward row vector
\[
v_{i-1}^{\left(\boldP_{:(i-1)}\right)}\in \mathbb{C}^{1\times \beta_{i-1}}
\]
has been constructed. For each candidate $Q\in \calP_1$, define
\[
\hat{v}_i^{(Q)}
:=
v_{i-1}^{\left(\boldP_{:(i-1)}\right)}\Gamma_i^{(Q)},
\qquad
\omega_i^{(Q)}
:=
\hat{v}_i^{(Q)}\hat{v}_i^{(Q)\dagger}.
\]
Because the unsampled right-canonical suffix contracts with its conjugate to the identity, we have
\begin{align}
\omega_i^{(Q)} & =
\sum_{\boldP_{>i}\in \calP_1^{n-i}}
\left|
\chi_{O_{\tilde{A}},\left(\boldP_{:(i-1)},Q,\boldP_{>i}\right)}
\right|^2\nonumber\\
& =
\sum_{\boldP_{>i}\in \calP_1^{n-i}}
\chi_{O_{\tilde{A}},\left(\boldP_{:(i-1)},Q,\boldP_{>i}\right)}^2.
\end{align}
Hence
\[
p(P_i=Q\mid \boldP_{<i})
=
\frac{\omega_i^{(Q)}}{\sum_{Q'\in \calP_1}\omega_i^{(Q')}}.
\]
We sample $P_i$ according to these weights and update
\[
v_i^{(\boldP_{:i})}:=\hat{v}_i^{(P_i)}.
\]
After the final step, $v_n^{(\boldP)}$ is a scalar and satisfies
\[
v_n^{(\boldP)}=\chi_{O_{\tilde{A}},\boldP}.
\]
In particular,
\[
Z=\sum_{Q\in \calP_1}\omega_1^{(Q)}.
\]
Therefore the same forward sweep simultaneously samples $\boldP$ from $\chi_{O_{\tilde{A}},\boldP}^2/Z$ and evaluates $\chi_{O_{\tilde{A}},\boldP}$. The corresponding non-grouped DFE random variable is
\[
R_\boldP=\frac{Z\,\chi_{\sigma,\boldP}}{\chi_{O_{\tilde{A}},\boldP}}.
\]

The grouped extension is obtained exactly as in Section~\ref{sec:mps_gdfe_derivation}. We again draw a sorting string
\[
\boldg\in\{X,Y,Z\}^n,
\]
form the representative string $\calR_\boldg(\boldP)$, and define
\[
p_\boldg(\boldP)
:=
\frac{1}{Z}
\sum_{\boldP'\in g(\calR_\boldg(\boldP))}
\chi_{O_{\tilde{A}},\boldP'}^2.
\]
The corresponding ideal grouped random variable is
\[
R_\boldP
=
\frac{\sum_{\boldP'\in g(\calR_\boldg(\boldP))}
\chi_{O_{\tilde{A}},\boldP'}\chi_{\sigma,\boldP'}}
{p_\boldg(\boldP)}.
\]

To compute $p_\boldg(\boldP)$ efficiently, keep the same preimage notation in Eq.~\eqref{eq:preimage_definition}.
Initialize
\[
H_0^{(\boldg,\boldP)}:=1,
\]
and recursively define, for $1\le i\le n$,
\begin{equation}
\label{eq:recursion_mpo_grouping}
H_i^{(\boldg,\boldP)}
:=
\sum_{Q\in r_{g_i}^{-1}(\calR_\boldg(\boldP)_i)}
\Gamma_i^{(Q)\dagger}H_{i-1}^{(\boldg,\boldP)}\Gamma_i^{(Q)}.
\end{equation}
By construction,
\[
H_n^{(\boldg,\boldP)}
=
\sum_{\boldP'\in g(\calR_\boldg(\boldP))}
\chi_{O_{\tilde{A}},\boldP'}^2
=
Z\,p_\boldg(\boldP).
\]
Also,
\[
|g(\calR_\boldg(\boldP))|
=
2^{|\{i:\calR_\boldg(\boldP)_i=g_i\}|}.
\]

After measuring $\sigma$ in the setting $\calR_\boldg(\boldP)$ and obtaining signs $\bolds\in\{\pm1\}^n$, we use the same local factors $M_i^{(\boldP,\bolds)}$ as in Section~\ref{sec:mps_gdfe_derivation} and define the snapshot estimator
\begin{equation}
\label{eq:sign_contraction_mpo_grouping}
\tilde{R}_{\boldP,\bolds}
=
\frac{1}{p_\boldg(\boldP)d}
\operatorname{tr}\!\left[
O_{\tilde{A}}
\bigotimes_{i=1}^n
M_i^{(\boldP,\bolds)}
\right].
\end{equation}
Its unbiasedness is proved exactly as in Appendix~\ref{section:mps_gdfe_snapshot_unbiased}, with $\rho_A$ replaced by $O_{\tilde{A}}$.

Similar to MPS-DFE, we set the shot allocation as
\[
\tilde{m}
=
\left\lceil
\frac{2|g(\calR_\boldg(\boldP))|\,Z}
{p_\boldg(\boldP)d l\epsilon^2}
\ln\frac{2}{\delta}
\right\rceil.
\]

% \begin{widetext}
\begin{table*}[t]
\centering
\caption{\justifying Summary of time complexities for MPS and MPO targets with and without grouping. Here, $n$ is the number of qubits, $l$ is the total number of measurement settings, and $N$ is the total number of shots amortized over $l$. $B_\mathrm{s}$ and $B_\mathrm{o}$ denote the maximum bond dimensions for the MPS and MPO, respectively.}
\label{tab:time_complexity_summary}
\begin{tabular}{cccccc}
\toprule
\multicolumn{1}{c}{\multirow{2}{*}{\textbf{Target}}} & \multicolumn{1}{c}{\multirow{2}{*}{\textbf{Grouping}}} & \multicolumn{3}{c}{\textbf{Offline}} & \multicolumn{1}{c}{\multirow{2}{*}{\textbf{Online}}} \\
\cmidrule(lr){3-5}%\cmidrule(lr){6-6}
& &
Preprocessing &
Sample settings &
Compute probs. &
\\
\midrule
\multicolumn{1}{c}{\multirow{2}{*}{MPS}} & \xmark  &
$O(n B_\mathrm{s}^{3})$ &
$O(l\,n B_\mathrm{s}^{3})$ &
-- &
$O(N\,n)$ \\
& \cmark &
$O(n B_\mathrm{s}^{3})$ &
$O(l\,n B_\mathrm{s}^{3})$ &
$O(l\,n B_\mathrm{s}^{5})$ &
$O(N\,n B_\mathrm{s}^{3})$ \\
\multicolumn{1}{c}{\multirow{2}{*}{MPO}} & \xmark  &
$O(n B_\mathrm{o}^{3})$ &
$O(l\,n B_\mathrm{o}^{2})$ &
-- &
$O(Nn)$ \\
& \cmark &
$O(n B_\mathrm{o}^{3})$ &
$O(l\,n B_\mathrm{o}^{2})$ &
$O(l\,n B_\mathrm{o}^{3})$ &
$O(N\,n B_\mathrm{o}^{2})$ \\
\bottomrule
\end{tabular}
\end{table*}
% \end{widetext}

\subsection{Time complexity}

In this section, we analyze the time complexity. Throughout, we use $l$ measurement settings and a total of $N$ shots, with the $N$ shots amortized over the $l$ settings. For the analysis, we assume that the bond dimension of the MPS $A$ is bounded by $B_\mathrm{s}$, and that the bond dimension of the MPO $\tilde A$ is bounded by $B_\mathrm{o}$. We consider four configurations: MPS versus MPO targets, and non-grouping versus grouping.

Across all four variants we split the classical workload into an \emph{offline} phase (everything that can be prepared before any copies of the unknown state $\sigma$ are measured) and an \emph{online} phase (estimation once measurement outcomes are observed). In particular, all Pauli-string sampling and any tensor-network objects that do not depend on the measurement outcomes can be built offline. Online work is only what remains after plugging in the observed signs. We summarize the resulting time complexities in Table~\ref{tab:time_complexity_summary}.

\paragraph{MPS, no grouping.}
Offline, we first gauge-transform the target MPS into right-canonical form with orthogonality center at the first tensor. This is a standard right-to-left orthonormalization sweep and costs $O(nB_\mathrm{s}^3)$. In this gauge, once the unsampled suffix is summed using Eq.~\eqref{equation:summation_as_swap}, the remaining right-canonical suffix contracts with its conjugate to the crossed identity (Fig.~\ref{fig:swap_simpler}), so no separate backward-contraction tensors need to be precomputed. To generate $l$ measurement settings, we sample each Pauli string sequentially. At site $i$, for each of the four candidates $P\in P_1$, we contract the current forward message $B_{i-1}^{\left(\boldP_{:(i-1)}\right)}$ with the local tensor $\tilde{B}_i^{(P)}$ from Eq.~\eqref{eq:B_tilde_definition}, evaluate the conditional weights $\beta_i^{(P)}$, sample $P_i$, and update the forward message. Using the factorized form on the right-hand side of Eq.~\eqref{eq:B_tilde_definition}, each site update costs $O(B_\mathrm{s}^3)$, so one full Pauli string costs $O(nB_\mathrm{s}^3)$ and all $l$ settings cost $O(lnB_\mathrm{s}^3)$. No separate ``Compute probs.'' stage is needed in the non-grouped MPS case. After the final site, the scalar forward message satisfies $B_{n,1,1}^{(\boldP)}=\mathrm{tr}(\rho_A \boldP)$, hence $\chi_{\rho_A,\boldP}=B_{n,1,1}^{(\boldP)}/\sqrt{d}$ is obtained as part of the same sampling sweep. Online, each shot returns $n$ local $\pm1$ outcomes in the sampled Pauli basis. Forming the product corresponding to $\boldP$ and updating the empirical estimate of $\chi_{\sigma,\boldP}$ therefore costs $O(n)$ per shot, giving a total online complexity of $O(Nn)$.

\paragraph{MPS, grouping.}
The precomputation and sampling of latent Pauli strings $\boldP$ are identical to the non-grouped case.
Grouping adds an additional offline step because each sampled latent $\boldP$ (together with a sorting string $\boldg$) induces a representative setting $\calR_\boldg(\boldP)$ and we must compute the group weight $p_\boldg(\boldP)$ to weight the grouped estimator and decide how many shots to allocate to that setting.
This $p_\boldg(\boldP)$ is computed by a sequential contraction similar to the MPS-DFE characteristic-function contraction, except that at certain sites we must sum over two possible single-qubit Paulis, which turns each site update into a contraction involving two $\tilde B$ tensors and the current rank-4 message $G$.
The right-hand side of Eq.~\eqref{eq:B_tilde_definition} allows each such site update to be implemented in $O(B_\mathrm{s}^5)$ time, giving $O(l n B_\mathrm{s}^5)$ overall.
Online, each shot produces a sign vector $\bolds$ and we evaluate the snapshot estimator. The snapshot trace is again an MPS expectation of a product of single-qubit operators (now $M_i^{(\boldP,\bolds)}$ depends on the observed sign at site $i$). Using the MPS-native contraction, each shot is $O(n B_\mathrm{s}^3)$, giving $O(N n B_\mathrm{s}^3)$ total online.

\paragraph{MPO, no grouping.}
Offline, we first contract each local MPO tensor $\tilde{A}_i$ with the single-qubit Pauli operator on its physical legs to form the induced operator-space matrices $\Gamma_i^{(P)}$, and gauge-transform the resulting MPS into right-canonical form with orthogonality center at the first tensor. This right-to-left orthonormalization sweep costs $O(nB_\mathrm{o}^3)$. We then sample each Pauli string by the forward recursion in Section~\ref{sec:mpo_algorithm}. At site $i$, for each of the four candidates $Q\in \calP_1$, we form the row vector $\hat{v}_i^{(Q)} = v_{i-1}^{\left(\boldP_{:(i-1)}\right)}\Gamma_i^{(Q)}$ and its weight $\omega_i^{(Q)}=\hat{v}_i^{(Q)}\hat{v}_i^{(Q)\dagger}$. Each candidate update costs $O(B_\mathrm{o}^2)$, so one full Pauli string costs $O(nB_\mathrm{o}^2)$, and all $l$ settings cost $O(lnB_\mathrm{o}^2)$. No separate ``Compute probs.'' stage is needed in the non-grouped MPO case because after the final site, the scalar forward message already equals $\chi_{O_{\tilde{A}},\boldP}$, and $Z$ is obtained from the same sweep. Online, each shot measures $n$ single-qubit Paulis and contributes the product of the observed signs to the estimator of $\chi_{\sigma,\boldP}$. Forming the corresponding single-shot DFE estimator therefore costs $O(n)$ per shot, giving a total online complexity of $O(Nn)$.

\paragraph{MPO, grouping.}
The preprocessing and latent-string sampling are the same as in the non-grouped MPO case, so the preprocessing cost is $O(nB_\mathrm{o}^3)$ and sampling $l$ latent strings costs $O(lnB_\mathrm{o}^2)$. Grouping adds an offline computation of the normalized group weight $p_\boldg(\boldP)$ for each sampled latent string. Using the recursion in Eq.~\eqref{eq:recursion_mpo_grouping}, each site update multiplies a $B_\mathrm{o}\times B_\mathrm{o}$ message on both sides by one or two $B_\mathrm{o}\times B_\mathrm{o}$ matrices, which costs $O(B_\mathrm{o}^3)$ per site. Hence computing all grouped probabilities costs $O(lnB_\mathrm{o}^3)$. Online, each shot yields a sign vector $\bolds$, and the grouped snapshot estimator requires the MPO contraction in Eq.~\eqref{eq:sign_contraction_mpo_grouping}. Contracting a single MPO with a product operator can be done by a standard boundary sweep in $O(nB_\mathrm{o}^2)$ time, so the total online cost is $O(NnB_\mathrm{o}^2)$.

\begin{figure}
    \centering
    \includegraphics[width=\linewidth]{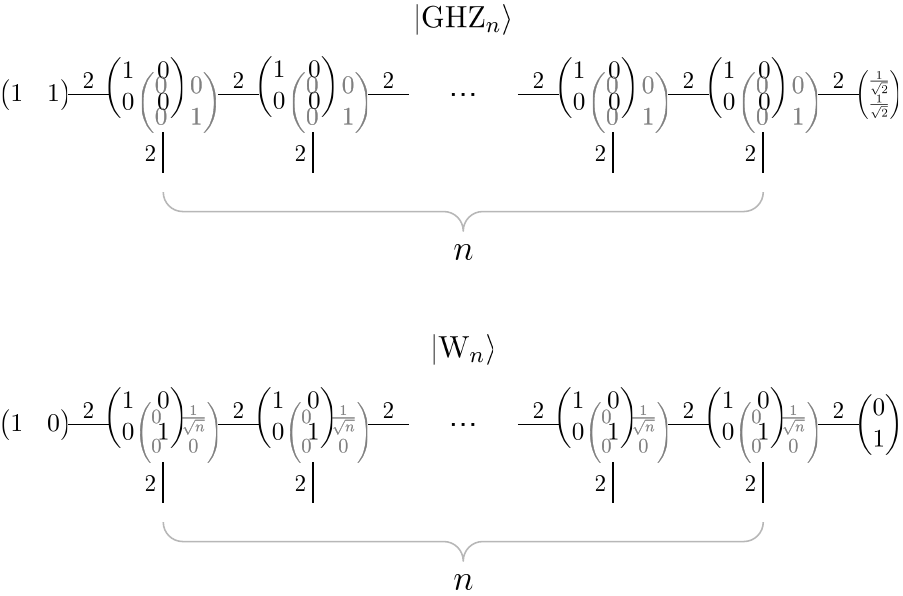}
    \caption{\justifying Both the GHZ state and the W state admit open-boundary MPS representations with bond dimension $2$. For each site, we display the two local matrices corresponding to the physical indices $b\in\{0,1\}$ (i.e., the local computational basis states).}
    \label{fig:ghz_w_state_mps}
\end{figure}

To obtain an overall polynomial-time algorithm, it is not enough for the \emph{per-shot} classical postprocessing to scale favorably. The \emph{total number of shots} $N$ must also remain manageable. In particular, while our per-shot computation is linear in $n$, the number of shots required per measurement setting must also avoid exponential scaling in order for the full fidelity-estimation pipeline to scale favorably with the number of qubits.
Our approach is most effective in regimes where the target MPS is ``well-conditioned'', a property signifying that most Pauli weights are not exponentially small \cite{flammia2011direct, cha2025efficient, sun2025efficient}.
For example, GHZ and W states satisfy this property, and they admit open-boundary MPS representations with bond dimension $2$ at all sites (see Fig.~\ref{fig:ghz_w_state_mps}). We do not benchmark these states here because specialized, more efficient algorithms are already established \cite{flammia2011direct, barbera2025sampling, cha2025efficient, cha2025operator}.

Even when the target MPS is not well-conditioned, MPS-DFE provides a near-quadratic speedup over the naive approach. Specifically, \emph{truncating bad events} yields a bound on the total number of shots in MPS-DFE of
$$
O\left(\frac{1}{\epsilon^2 \delta} + \frac{d\log(1/\delta)}{\epsilon^2}\right),
$$
at the cost of introducing a small bias in the fidelity estimate~\cite{flammia2011direct}. Consequently, the total classical compute time scales as $O(d\log d)$ (suppressing additional dependence on precision parameters and bond dimension). By contrast, the naive approach requires at least $\Omega(d^2 \log d)$ time, since it must evaluate $\chi_{\rho_A,\boldP}$ for all $d^2$ Pauli operators, and even with efficient tensor contractions each evaluation costs $\Omega(\log d)$.

\begin{figure}[t]
  \centering
  \includegraphics[width=\linewidth]{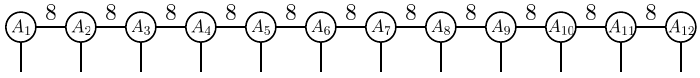}
  \caption{\justifying The $n=12$ open-boundary random MPS target used in our numerical simulations (before canonicalization), based on the construction in \cite{malz2024preparation}. The bond dimensions are annotated above each virtual bond.}
  \label{fig:experiment_MPS}
\end{figure}

\begin{figure}[h]
     \centering
     
     \begin{subfigure}[b]{\linewidth}
         \centering
         \includegraphics[width=\linewidth]{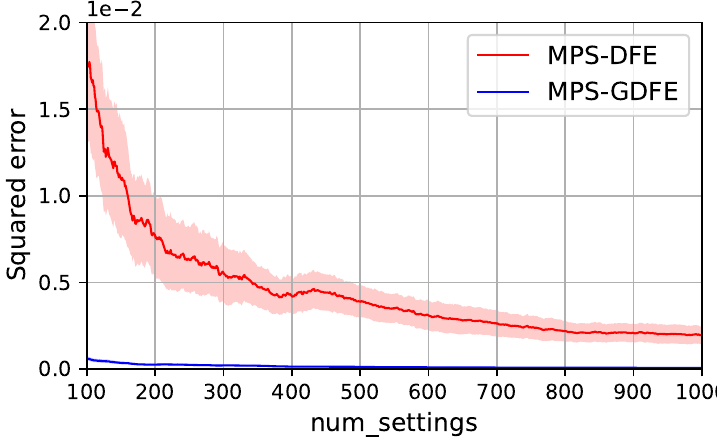}
         \caption{}
         \label{fig:result_MSE}
     \end{subfigure}
     \begin{subfigure}[b]{\linewidth}
         \centering
         \includegraphics[width=\linewidth]{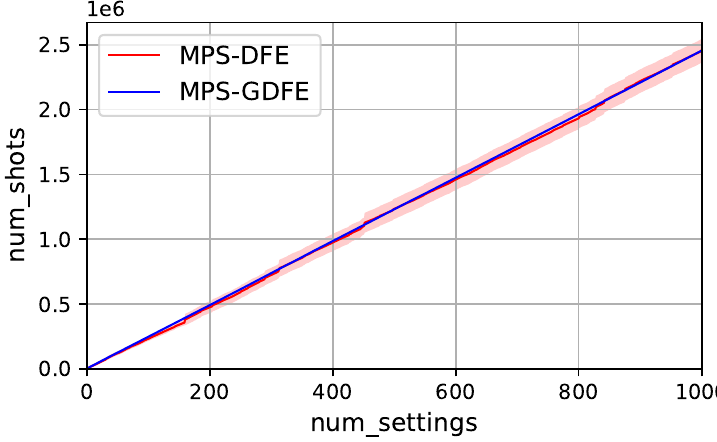}
         \caption{}
         \label{fig:result_shots}
     \end{subfigure}
     
     \caption{\justifying Numerical results for MPS-DFE and MPS-GDFE on the $n=12$ random MPS target. (a) Mean squared error (MSE) and (b) total number of shots for MPS-DFE and MPS-GDFE, averaged over 100 independent trials. Shaded regions denote 95\% confidence intervals.}
     \label{fig:combined_results}
\end{figure}

\section{Numerical results}

Numerical simulations were performed on an $n=12$ random MPS with open boundaries, following the construction in \cite{malz2024preparation} and illustrated in Fig.~\ref{fig:experiment_MPS}.
Each local tensor was generated at random, and the resulting state was normalized to ensure $\langle\psi_A|\psi_A\rangle=1$.
For the noisy simulations, we consider a depolarized state
$$
\sigma = (1-\lambda)|\psi_A\rangle\langle\psi_A| + \lambda \frac{I}{d}
$$
with a noise parameter of $\lambda = 0.1$.
Using this target instance, we compared our non-grouped estimator (MPS-DFE) with its grouped variant (MPS-GDFE).

We set $\epsilon = \delta = 0.1$, resulting in $l = 1000$. Fig.~\ref{fig:result_MSE} summarizes the results.
The online estimator converges to the ground-truth
fidelity as the number of sampled measurement settings increases, and grouping yields a substantial variance reduction.
Averaging over $100$ independent trials, MPS-GDFE achieves a 97\% reduction in MSE after $l$ measurement settings, while requiring essentially the same total number of shots as MPS-DFE.

In contrast to the slight copy reduction reported for grouped DFE in~\cite{barbera2025sampling}, Fig.~\ref{fig:result_shots} demonstrates that our total shot counts do not decrease. This is because we replace the $\ell_1$-mass term used for shot allocation with a computationally convenient $\ell_2$-based upper bound.

\section{Discussion}

We take a step toward making matrix product states verifiable in the NISQ era.
While much prior work emphasizes preparing MPS with shallow circuits, scalable certification is what turns preparation into a trustworthy practical benchmark and validates conclusions drawn from quantum simulation.
Using only local Pauli measurements, our protocol leverages the target MPS to sample Pauli strings sequentially from the correct importance distribution, eliminating the exponential preprocessing bottleneck of standard DFE and reducing sampling overhead to linear in $n$.
A grouped variant further reuses each single-qubit setting to estimate an entire qubit-wise commuting group, suppressing estimator variance without sacrificing efficient postprocessing, and the same ideas extend naturally to MPO targets.

Several directions could broaden scalability and scope.
Although per-shot postprocessing is linear in system size, the remaining contractions still scale polynomially with bond dimension.
Tightening these dependencies (or introducing controlled approximations with provable bias/variance tradeoffs) would make longer, more entangled chains accessible.
Beyond Pauli sampling, recent approaches optimize an operator expansion over an overcomplete $6^n$-element POVM \cite{acharya2021shadow, fischer2024dual, jameson2024optimal, cha2025operator}.
Adapting that optimization to tensor-network contractions may yield new samplers for MPS.
Finally, moving beyond Pauli measurements, combining MPS-friendly postprocessing with shallow-shadow measurement ensembles is a promising route to richer observables and improved noise robustness \cite{cioli2025approximate, farias2025robust, hu2025demonstration, zhang2025holographic}.
% \color{black}

% \begin{acknowledgments}
% \TODO
% \end{acknowledgments}

% The \nocite command causes all entries in a bibliography to be printed out
% whether or not they are actually referenced in the text. This is appropriate
% for the sample file to show the different styles of references, but authors
% most likely will not want to use it.
% \nocite{*}
\bibliography{apssampv1}% Produces the bibliography via BibTeX.

% \clearpage
% \appendix

% SUPP REF
% \putbib[apssampv1]
% \end{bibunit}

\clearpage
\onecolumngrid
\appendix

% \begin{center}
% \textbf{Supplemental Material: Verifying matrix product states with autoregressive local measurements}
% \end{center}

% \setcounter{section}{0}
% \setcounter{figure}{0}
% \setcounter{table}{0}
% \setcounter{equation}{0}

% --- SUPPLEMENT LABELS: prefix S ---
% \renewcommand{\thesection}{S\arabic{section}}
% \renewcommand{\thefigure}{S\arabic{figure}}
% \renewcommand{\thetable}{S\arabic{table}}
% \renewcommand{\theequation}{S\arabic{equation}}
% \renewcommand{\thesubsection}{\thesection.\arabic{subsection}}
% \renewcommand{\thesubsubsection}{\thesubsection.\arabic{subsubsection}}
% \makeatletter
% \renewcommand{\fnum@figure}{Fig.~\thefigure}
% \renewcommand{\fnum@table}{Table~\thetable}
% \makeatother

% SUPP REF
% \begin{bibunit}

\section{Tensor networks}
\label{sec:tensor_networks}

\begin{figure}[h]
    \centering
    \begin{tikzpicture}[scale=1.0, baseline=(current bounding box.center)]
      \tikzset{
        tensor/.style={draw, rounded corners, minimum width=9mm, minimum height=6mm, inner sep=1pt},
        leg/.style={line width=0.5pt},
      }

      % Two tensors A and B with a contracted bond b and open legs a,c,i,j
      \node[tensor] (A) at (0,0) {$A$};
      \node[tensor] (B) at (2.2,0) {$B$};

      % contracted edge b
      \draw[leg] (A.east) -- node[above] {$b$} (B.west);

      % open legs for A
      \draw[leg] (A.west) -- ++(-0.9,0) node[left] {$a$};
      \draw[leg] (A.north) -- ++(0,0.9) node[above] {$i$};

      % open legs for B
      \draw[leg] (B.east) -- ++(0.9,0) node[right] {$c$};
      \draw[leg] (B.north) -- ++(0,0.9) node[above] {$j$};

      % Result label
      \node at (1.1,-0.9) {$C_{acij}=\sum_b A_{abi}B_{bcj}$};
    \end{tikzpicture}
    \caption{A single contraction. Connecting an edge corresponds to summing over the associated index.}
    \label{fig:tn_basic_contraction}
\end{figure}
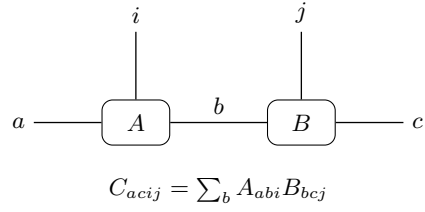

Tensor networks provide a compact way to represent and manipulate high-dimensional vectors and operators by decomposing them into \emph{local} tensors connected by contracted indices.
A (complex) \emph{tensor} is simply a multidimensional array
$T = (T_{i_1,i_2,\dots,i_r})$ with $r$ indices, where each index $i_j$ ranges over a finite set of size $\dim(i_j)$.
\emph{Tensor product} corresponds to placing tensors side-by-side without summing any indices, and
\emph{tensor contraction} corresponds to summing over one or more shared indices.

\paragraph{Graphical notation.}
We use the standard graphical calculus in which each tensor is drawn as a node and each index is drawn as an edge attached to that node.
An \emph{open} edge represents an uncontracted (free) index, while an edge connecting two nodes represents a \emph{contracted} index that is summed over.
A diagram with no open edges evaluates to a scalar.
A convenient mnemonic is:
\[
\text{number of open edges} \quad \leftrightarrow \quad \text{tensor order of the resulting object.}
\]
For example, a vector has one open edge, a matrix has two, and a scalar has none.

\paragraph{Contraction.}
If two tensors share an index, then connecting the corresponding edges denotes summation over that index.
For instance, given $A_{ab i}$ and $B_{bc j}$, contracting the shared index $b$ produces a tensor $C_{acij}$:
\begin{equation}
\label{eq:basic_contraction}
C_{a c i j} \;=\; \sum_{b} A_{a b i}\, B_{b c j}.
\end{equation}
In Fig.~\ref{fig:tn_basic_contraction}, the summed index is the internal edge.

\paragraph{Gauge freedom, right-canonical form, and orthogonality center.}
For each site \(i\) and physical value \(b_i\in\{0,1\}\), we denote by
\[
A_i^{(b_i)} \in \mathbb{C}^{\alpha_{i-1}\times \alpha_i}
\]
the matrix slice of the local tensor \(A_i\), defined by
\[
\bigl(A_i^{(b_i)}\bigr)_{a_{i-1},a_i}
:= A_{i,a_{i-1},a_i,b_i}.
\]
Then the open-boundary MPS can be written compactly as
\[
|\psi_A\rangle
=
\sum_{b\in\{0,1\}^n}
\bigl(A_1^{(b_1)}A_2^{(b_2)}\cdots A_n^{(b_n)}\bigr)_{1,1}
\,|b\rangle.
\]

This representation is not unique. Indeed, for any collection of invertible matrices
\(X_i\in\mathbb{C}^{\alpha_i\times \alpha_i}\) (\(i=1,\dots,n-1\)),
the gauge transformation
\[
A_i^{(b_i)} \mapsto A_i^{(b_i)}X_i,
\qquad
A_{i+1}^{(b_{i+1})} \mapsto X_i^{-1}A_{i+1}^{(b_{i+1})}
\]
leaves the physical state unchanged.

Canonical forms are particular gauge choices that
make the virtual indices orthonormal. Specifically, an open-boundary MPS is in mixed-canonical form with orthogonality center
at site \(k\) if the tensors to the left of \(k\) are left-canonical,
\[
\sum_{b_i=0}^1 A_i^{(b_i)\dagger}A_i^{(b_i)} = I_{\alpha_i},
\qquad i<k,
\]
and the tensors to the right of \(k\) are right-canonical,
\[
\sum_{b_i=0}^1 A_i^{(b_i)}A_i^{(b_i)\dagger} = I_{\alpha_{i-1}},
\qquad i>k.
\]
The tensor at site \(k\) is then called the orthogonality center.

The special case relevant here is \(k=1\). For brevity, we refer to this as a
right-canonical MPS with orthogonality center at the first tensor. Concretely, this means
that all tensors \(A_i\) with \(i=2,\dots,n\) satisfy
\[
\sum_{b_i=0}^1 A_i^{(b_i)}A_i^{(b_i)\dagger}
=
I_{\alpha_{i-1}},
\qquad i=2,\dots,n.
\]
In index notation, the same condition reads
\[
\sum_{b_i=0}^1 \sum_{a_i=1}^{\alpha_i}
A_{i,a_{i-1},a_i,b_i}
A_{i,a'_{i-1},a_i,b_i}^{*}
=
\delta_{a_{i-1},a'_{i-1}},
\qquad i=2,\dots,n.
\]

A useful consequence is that contracting any right-canonical suffix with its conjugate collapses to the
identity on the incoming bond:
\[
\sum_{b_i,\dots,b_n}
\bigl(A_i^{(b_i)}\cdots A_n^{(b_n)}\bigr)
\bigl(A_i^{(b_i)}\cdots A_n^{(b_n)}\bigr)^{\dagger}
=
I_{\alpha_{i-1}},
\qquad i=2,\dots,n.
\]
In particular, for a normalized state and \(\alpha_0=1\),
\begin{align*}
\langle \psi_A | \psi_A\rangle
& =
\sum_{b\in\{0,1\}^n}
\bigl(A_1^{(b_1)}A_2^{(b_2)}\cdots A_n^{(b_n)}\bigr)_{1,1}
\bigl(A_1^{(b_1)}A_2^{(b_2)}\cdots A_n^{(b_n)}\bigr)_{1,1}^\ast\\
& =
\sum_{b_1=0}^1
A_1^{(b_1)}
\left( \sum_{b_2,\dots,b_n}
\bigl(A_2^{(b_2)}\cdots A_n^{(b_n)}\bigr)
\bigl(A_2^{(b_2)}\cdots A_n^{(b_n)}\bigr)^{\dagger} \right)
A_1^{(b_1)\dagger}\\
& = \sum_{b_1=0}^1
A_1^{(b_1)}A_1^{(b_1)\dagger}\\
& = 1.
\end{align*}

Finally, any open-boundary MPS can be brought into this form by a right-to-left sequence
of gauge transformations (equivalently, by a QR- or SVD-based orthonormalization sweep).
This changes only the tensor representation and not the physical state.

\section{Sampling groups of Pauli operators}
\label{subsec:grouping-pauli}

A well-established way to reduce the variance of a fidelity estimator is to group Pauli strings that can be measured simultaneously \cite{barbera2025sampling}.

\paragraph{Qubit-wise commutativity.}
Recall that an $n$-qubit Pauli string is
$\boldP = P_1\otimes \cdots \otimes P_n \in \calP_1^{\otimes n}$.
Two Pauli strings $\boldP$ and $\boldP^\prime$ are said to be \emph{qubit-wise commuting (QWC)} if they commute on every qubit,
i.e.,
\begin{equation}
\label{eq:qwc-def}
[P_i, P^\prime_i]=0 \quad \text{for all } i=1,\dots,n.
\end{equation}
Equivalently, for each qubit $i$, either $P_i=I$, or $P^\prime_i=I$, or $P_i=P^\prime_i$.

\paragraph{Group weights and shots per setting.}
Let the target state be $\rho$ and define the characteristic function
$\chi_{\rho,\boldP}=\tr(\rho \boldP)/\sqrt d$.
Given a QWC group $g\subseteq\calP_1^n$, define its group weight and $\ell_1$-mass as
\begin{equation}
\label{eq:group-weight}
p(g) := \sum_{\boldP\in g} \chi_{\rho,\boldP}^2,
\qquad
\|\chi_{\rho}\|_{1,g} := \sum_{\boldP\in g} |\chi_{\rho,\boldP}|.
\end{equation}
In grouped DFE (GDFE), one samples measurement settings (groups) according to $p(g)$, and then estimates the
corresponding grouped contribution using repeated measurements in that setting.
Throughout the paper, we adopt the \emph{non-overlapping} framework, in which the groups form a partition of the Pauli strings $\calP_1^{\otimes n}$.
Given the sampled group $g$, define the \emph{ideal} (infinite-shot) group estimator
\begin{equation}
\label{eq:group_rv_ideal}
R_g
:=
\frac{1}{p(g)}\sum_{\boldP\in g} \chi_{\rho,\boldP} \chi_{\sigma,\boldP}.
\end{equation}
Then $R_g$ is an unbiased estimator of $\tr(\rho\sigma)$ under the sampling distribution $p(g)$:
\begin{align}
\mathbb{E}_g[R_g] =
\sum_{g} p(g) \frac{1}{p(g)}\sum_{\boldP\in g}\chi_{\rho,\boldP}\chi_{\sigma,\boldP}
=
\sum_{\boldP}\chi_{\rho,\boldP}\chi_{\sigma,\boldP}
=
\tr(\rho\sigma).
\end{align}
In practice, $R_g$ is unknown and must be estimated from repeated measurements of the \emph{single} measurement setting associated with the QWC group $g$.
A convenient Hoeffding-type choice for the number of copies allocated to a sampled group $g$ is
\begin{equation}
\label{eq:shots_per_setting}
m
=
\left\lceil
\frac{2\|\chi_{\rho}\|_{1,g}^2}{p(g)^2 d l \epsilon^2}
\ln\!\frac{2}{\delta}
\right\rceil,
\end{equation}
where $l$ is the number of i.i.d. group-samples in the outer Monte Carlo loop (typically $l=\lceil 1/(\epsilon^2\delta)\rceil$).

\section{Evaluation of conditional Pauli distribution}
\label{section:detailed_conditional_swap_equation}

It follows from Eqs.~\eqref{equation:prob_as_double_copy}~and~\eqref{equation:summation_as_swap} that for $2\le i\le n$,
\begin{align*}
% \label{eq:swap_formula}
\sum_{\boldP:\,\boldP_{<i}=\boldP^{(\mathrm{l})}} p(\boldP)
% & = \frac{1}{d} \sum_{\boldP^{(\mathrm{h})}\in\calP_1^{\otimes (n-i+1)}} \tr\!\left(\rho_A^{\otimes2}\left(\boldP_\mathrm{l}^{\otimes2} \otimes \boldP_\mathrm{h}^{\otimes2}\right)\right)\nonumber\\
% & = \frac{1}{d} \tr\!\left(\rho_A^{\otimes2}\left(\boldP_\mathrm{l}^{\otimes2} \otimes \sum_{\boldP_\mathrm{h}\in\calP_1^{\otimes (n-i+1)}} \boldP_\mathrm{h}^{\otimes2}\right)\right)\\
% & = \frac{1}{d} \tr\!\left(\rho_A^{\otimes2}\left(\boldP_\mathrm{l}^{\otimes2} \otimes S_{n-i+1}^\dagger (2\times \mathrm{SWAP})^{\otimes (n-i+1)} S_{n-i+1}\right)\right)\\
& = \frac{1}{d} \tr\!\left( \tilde\rho^{(2)} \left( \bigotimes_{j<i} P^{(\mathrm{l})\otimes2}_j \otimes \sum_{\boldP^{(\mathrm{h})}\in\calP_1^{n-i+1}} \bigotimes_{j\ge i} P^{(\mathrm{h})\otimes2}_{j-i+1} \right) \right)\\
& = \frac{1}{d} \tr\!\left( \tilde\rho^{(2)} \left( \bigotimes_{j<i} P^{(\mathrm{l})\otimes2}_j \otimes (2\times \mathrm{SWAP})^{\otimes(n-i+1)} \right) \right)\\
& = \frac{1}{d} \tr\!\left( \rho_A^{\otimes2} S_n^\dagger \left( \bigotimes_{j<i} P^{(\mathrm{l})\otimes2}_j \otimes (2\times \mathrm{SWAP})^{\otimes(n-i+1)} \right) S_n \right),
\end{align*}
which can be used to evaluate both the numerator and denominator in Eq.~\eqref{eq:conditional_formula} efficiently.

\section{Unbiasedness of MPS-GDFE snapshot}
\label{section:mps_gdfe_snapshot_unbiased}

Fix $\boldP$ and the sampled sorting string $\boldg$, and let $\boldQ := R_\boldg(\boldP)$ denote the actually measured Pauli string.
Let $\bolds\in\{\pm1\}^n$ be the measurement outcome, and let
\[
\Pi^{(\boldQ)}_{\bolds} := \bigotimes_{i=1}^n \frac{I + s_i Q_i}{2},
\qquad
p(\bolds \mid \boldQ) = \tr\!\left(\sigma\Pi^{(\boldQ)}_{\bolds}\right).
\]
Recall
\[
\tilde R_{\boldP,\bolds}
=
\frac{1}{p_\boldg(\boldP)d}
\tr\!\left(\rho_A \bigotimes_{i=1}^n M^{(\boldP,\bolds)}_i\right).
\]
First, expand the tensor product as
\begin{equation}
\label{equation:M_i_product_expansion}
\bigotimes_{i=1}^n M^{(\boldP,\bolds)}_i
=
\sum_{\boldP^\prime \in g(\boldQ)} \eta_\bolds(\boldP^\prime)\boldP^\prime,
\qquad
\eta_\bolds(\boldP^\prime) := \prod_{i:P^\prime_i\neq I} s_i.
\end{equation}
Then, by definition of expectation,
\begin{align*}
\mathbb{E}_\bolds\!\left[\tilde R_{\boldP,\bolds}\right]
&=
\sum_{\bolds\in\{\pm1\}^n} p(\bolds\mid \boldQ)\tilde R_{\boldP,\bolds}
\\
&=
\frac{1}{p_\boldg(\boldP)d}
\sum_{\bolds\in\{\pm1\}^n}
\tr\!\left(\sigma \Pi^{(\boldQ)}_\bolds\right)
\tr\!\left(\rho_A \bigotimes_{i=1}^n M^{(\boldP,\bolds)}_i\right).
\end{align*}
Substitute the expansion Eq.~\eqref{equation:M_i_product_expansion} and swap the order of summation:
\begin{align}
\label{equation:expectation_long_expansion}
\mathbb{E}_\bolds\!\left[\tilde R_{\boldP,\bolds}\right]
&=
\frac{1}{p_\boldg(\boldP)d}
\sum_\bolds
\tr\!\left(\sigma \Pi^{(\boldQ)}_\bolds\right)
\tr\!\left(\rho_A \sum_{\boldP^\prime \in g(\boldQ)} \eta_\bolds(\boldP^\prime)\boldP^\prime \right) \nonumber
\\
&=
\frac{1}{p_\boldg(\boldP)d}
\sum_{\boldP^\prime \in g(\boldQ)}
\tr(\rho_A \boldP^\prime)
\sum_\bolds
\eta_\bolds(\boldP^\prime) \tr\!\left(\sigma \Pi^{(\boldQ)}_\bolds\right).
\end{align}
Next we evaluate the inner sum. For any fixed $\boldP^\prime \in g(\boldQ)$, $\boldP^\prime$ is diagonal in the joint eigenbasis of $\boldQ$,
and its eigenvalue on the projector $\Pi^{(\boldQ)}_{\bolds}$ is exactly $\eta_\bolds(\boldP^\prime)$. Equivalently,
\[
\Pi^{(\boldQ)}_{\bolds} \boldP^\prime = \eta_\bolds(\boldP^\prime) \Pi^{(\boldQ)}_{\bolds}.
\]
Summing over $\bolds$ gives the spectral decomposition
\[
\boldP^\prime = \sum_{\bolds} \eta_\bolds(\boldP^\prime) \Pi^{(\boldQ)}_{\bolds}.
\]
Therefore,
\begin{align}
\label{equation:eta_sum_substitute}
\sum_\bolds \eta_\bolds(\boldP^\prime) \tr\!\left(\sigma \Pi^{(\boldQ)}_\bolds\right)
&=
\tr\!\left(\sigma \sum_{\bolds}\eta_\bolds(\boldP^\prime) \Pi^{(\boldQ)}_{\bolds}\right) \nonumber \\
&=
\tr(\sigma \boldP^\prime).
\end{align}
Plugging Eq.~\eqref{equation:eta_sum_substitute} into Eq.~\eqref{equation:expectation_long_expansion} yields
\[
\mathbb{E}_\bolds\!\left[\tilde R_{\boldP,\bolds}\right]
=
\frac{1}{p_\boldg(\boldP)d}
\sum_{\boldP^\prime\in g(\boldQ)}
\tr(\rho_A \boldP^\prime) \tr(\sigma \boldP^\prime).
\]
Finally, using $\chi_{A,\boldP^\prime} := \tr(A \boldP^\prime)/\sqrt{d}$,
we have $\tr(\rho_A \boldP^\prime)\tr(\sigma \boldP^\prime) = d\chi_{\rho_A,\boldP^\prime}\chi_{\sigma,\boldP^\prime}$,
so
\[
\mathbb{E}_\bolds\!\left[\tilde R_{\boldP,\bolds}\right]
=
\frac{1}{p_\boldg(\boldP)}
\sum_{\boldP^\prime \in g(\boldQ)}
\chi_{\rho_A,\boldP^\prime}\chi_{\sigma,\boldP^\prime}
=
R_\boldP.
\]

% SUPP REF
% \putbib[apssampv1]
% \end{bibunit}

\end{document}